\documentclass[12pt]{iopart}
\usepackage{graphicx}
\usepackage{amssymb}
\newcommand{\text}[1]{\hbox{\scriptsize\rm #1}}
\newcommand{\PRD}{{\it Phys. Rev.} D}

\begin{document}
\title[LIGO Inspiral Hardware Injections]
{Testing the LIGO Inspiral Analysis with Hardware Injections}
\author{Duncan A. Brown$^1$ for the LIGO Scientific Collaboration}
\address{$^1$ Department of Physics, University of Wisconsin--Milwaukee,
  Milwaukee, Wisconsin 53211, USA}
\begin{abstract}
  Injection of simulated binary inspiral signals into detector hardware
  provides an excellent test of the inspiral detection pipeline.  By
  recovering the physical parameters of an injected signal,  we test our
  understanding of both instrumental calibration and the data analysis
  pipeline. We describe an inspiral search code and results from
  hardware injection tests and demonstrate that injected signals can be
  recovered by the data analysis pipeline. The parameters of the recovered
  signals match those of the injected signals.
\end{abstract}
\pacs{04.80.Nn, 07.05.Kf, 97.80.--d, 01.30.Cc}
\ead{duncan@gravity.phys.uwm.edu}
\maketitle

\section{Introduction}
\label{s:intro}
Gravitational radiation incident on the LIGO interferometers from an
inspiralling binary will cause the test masses to move relative to each other.
This produces a differential change in length of the arms\cite{saulson}.
\emph{Injection} is the process of adding a waveform to interferometer data to
simulate the presence of a signal in the noise. We use injections to measure
the performance of the binary inspiral analysis pipeline\cite{abbott2003b}.
\emph{Software injections}, which add a simulated signal to the data after it
has been recorded, are used for efficiency measurements. Since this they
performed \emph{a posteriori} interferometer is not affected while it is
recording data.  Alternatively, a simulated signal can be added to the
interferometer control system to make the instrument behave as if an inspiral
signal is present. We call this \emph{hardware injection}; the data recorded
from the instrument contains the simulated signal.

Analysis of hardware injections ensures that the analysis pipeline is
sensitive to real inspiral signals and validates the software injections used
to test the pipeline efficiency.  In order to perform an accurate upper limit
analysis for binary neutron stars, we must measure the efficiency of our
pipeline\cite{abbott2003b}. That is, we inject a known number of signals into
the pipeline and determine the fraction of these detected.  Injecting signals
into the interferometer for the duration of a run is not practical, so we use
the analysis software to inject inspiral signals in software.  By comparing
software and hardware injections we confirm that software injections are
adequate to measure the efficiency of the upper limit pipeline.

Hardware injections provide a very complete method of testing the inspiral
detection pipeline. By recovering the physical parameters of an injected
signal, we test our understanding of all aspects of the pipeline, including
the instrument calibration, filtering algorithm and veto safety. We injected
inspiral signals immediately after the first LIGO science run (S1) in
August---September 2002. The data taken during this time was analyzed using
with the software tools used to search for real signals.  This article
describes the results from hardware injection tests performed by the Inspiral
Working Group of the LIGO Scientific Collaboration (LSC).

\section{Injection of the Inspiral Signals}
\label{s:injecting}

To inject the signals, we generate the interferometer strain, $h(t)$, produced
by an inspiralling binary using the restricted second order post-Newtonian
approximation in the time domain\cite{biww}.  The LSC calibration group
supplies a transfer function, $T(f)$, which allows us to construct a signal,
$v(t)$, that produces the desired strain when it is injected into the
interferometer.  The transfer function, $T(f)$, is given by \begin{equation}
T(f) = \frac{L}{C}\frac{f^2}{f_0^2}
\end{equation}
where $L$ is the length of the interferometer, $C$ is the calibration of the
excitation point in nm/count and $f_0$ is the pendulum frequency of the test
mass. Damping is neglected as it is unimportant in the LIGO frequency band.
Codes are available in the LIGO Algorithm Library (LAL)\cite{lal} for
simulating inspiral waveforms for hardware injection.

The interferometer \emph{Length Sensing and Control} system\cite{abbott2003a}
has excitation points which allow arbitrary signals to be added into the servo
control loops or to the drives that control the motion of the
mirrors\cite{shawhan2002}. 
During S1, we injected signals corresponding to an optimally oriented binary.
Injections of two $1.4\,M_\odot$ neutron stars at distances from $10$ kpc to
$80$ kpc were used to test the neutron star analysis. 
We also injected signals from a $1.4,\,4.0\,M_\odot$ binary and several
$1.4,1.4\,M_\odot$ binaries at closer distances.  These signals were injected
into the differential mode servo and directly into an end test mass drive.

\section{The S1 Inspiral Analysis Pipeline}
\label{ss:pipeline}

The data analysis pipeline used in the neutron star inspiral search is
described in \cite{abbott2003b}. In brief, we use matched filtering with a
bank of templates between $1.0$ and $3.0$ $M_\odot$ for each element of the
binary. This generates a list of triggers which exceed a signal-to-noise
threshold and pass a waveform quality test, known as the $\chi^2$ test, which
has proved effective at excluding triggers that are due to glitches in the
data. Auxiliary interferometer channels are filtered for glitches and those
inspiral triggers that are coincident with a glitch are vetoed.  We test for
coincident triggers, subject to the less sensitive interferometer being able
to see the trigger. Triggers that pass all cuts are considered events.  The
efficiency of the pipeline is measured by a Monte-Carlo simulation which uses
software injections into interferometer data.

\section{Results}

\subsection{Detection of the Injected Signals}
\label{ss:detection}

Table \ref{t:triggers} shows the events generated by processing 4000 seconds
of data from the Livingston 4 km interferometer (L1) on 10 September 2002
during the post-run hardware injections.  It can be seen that all of
the hardware injections are identified as candidate events since they have
high signal-to-noise ratios and values of the $\chi^2$ test lower than $5$,
which was the threshold used in the S1 analysis pipeline\cite{abbott2003b}.
\begin{table}[htb]
  \begin{flushright}
  \begin{tabular}{l|l|c|c}
  End time of Injection&End Time of Detection&$\rho$&$\chi^2$\\
  \hline
  $04:35:12.424928$ & $04:35:12.424927$ & $11.623546$ & $1.653222$ \\
  $04:36:42.424928$ & $04:36:42.425171$ & $20.230101$ & $1.671016$ \\
  $04:38:12.424928$ & $04:38:12.424927$ & $37.488770$ & $0.443966$ \\
  $04:39:42.424928$ & $04:39:42.424927$ & $69.815262$ & $1.375486$ \\
  \end{tabular}
  \end{flushright}
  \caption{%
  Hardware injection events found by the inspiral analysis pipeline. End time
  of injection is the known end time of the injected signal and end time of
  detection is the end time of the signal as reported by the analysis
  pipeline. Times are Universal Time (UTC) on 10 September 2002. The values of
  signal-to-noise ratio $\rho$ and $\chi^2$ veto are given for each event.
  }
\label{t:triggers}
\end{table}

Since we know the exact end times of the injected signals, we can compare them
with the value reported by the search code and ensure that the pipeline is
reporting the correct time. The raw data is resampled to $4096$ Hz before
being filtered. For each of the signals injected, we were able to detect the
coalescence time of the injection to within one sample point of the correct
value at $4096$ Hz, which is consistent with the expected statistical error
and confirms that the pipeline has not introduced any distortion of the
signals.

\subsection{Checking the Instrumental Calibration}
\label{s:calibration}

Calibration measurements of the interferometers were performed before and
after the run; these are the reference calibrations. In general, the
calibration changes due to changes in the alignment on timescales of minutes.
This variation can be encoded in a single parameter $\alpha$ which is
monitored using a sinusoidal signal injected into the
detector (the calibration line)\cite{adhikari2003}. $\alpha$ is used as input
to the data analysis pipeline and varied between $0.4$ and $1.4$ during S1.
Data is analysed in 256 second segments.  For each 256 seconds of data
starting at time $t_0$, we construct the calibration, $R(f;t_0)$ by using
$\alpha(t_0)$ and a reference calibration.  $R(f;t_0)$ is then used to
calibrate 256 seconds of data.

\begin{figure}[htb]
  \vspace{5pt}
  \begin{flushright}
    \includegraphics[width=\textwidth]{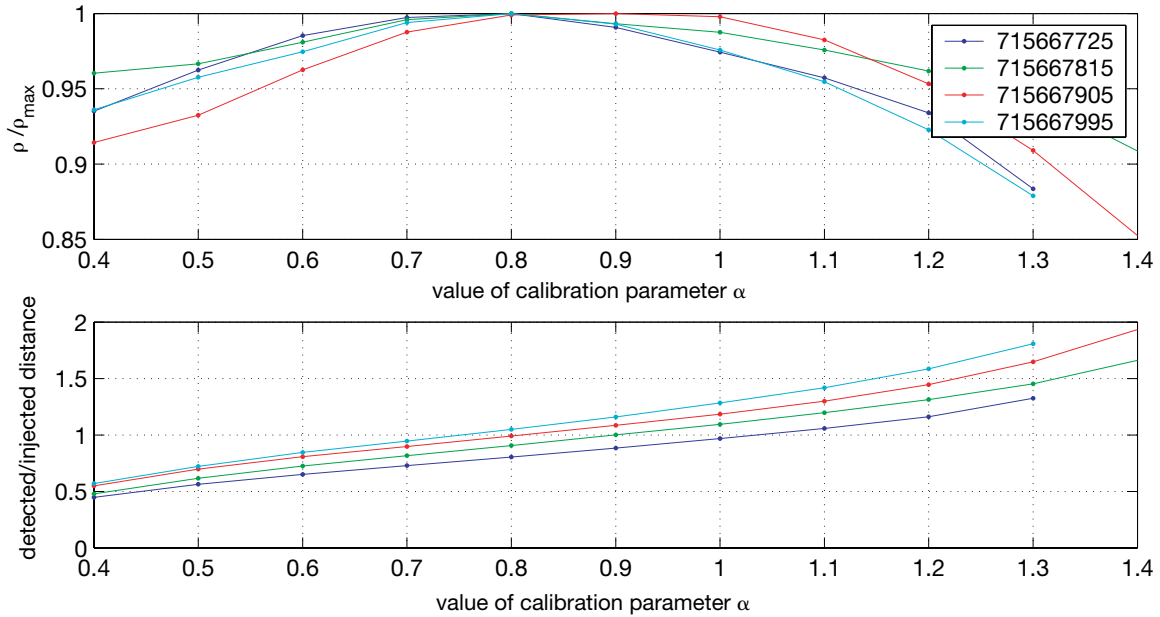}    
  \end{flushright}
  \caption{%
  Each curve corresponds to a hardware injection at the given GPS time. We
  re-analyse each injection with different calibrations to show how the
  detected quantities vary with $\alpha$. The upper plot shows the ratio of
  signal-to-noise ratio, $\rho$, to its maximum value, $\rho_{\mathrm{max}}$.
  The lower plot shows ratio of the detected distance to the known distance of
  the hardware injection.
  }
\label{f:calibration}
\end{figure}
Figure \ref{f:calibration} shows a set of injections into the Livingston
interferometer analyzed with different calibrations generated by varying the
value of $\alpha$. We expect that the signal-to-noise varies quadratically and
the effective distance varies linearly with changes in $\alpha$\cite{bruce}.
This is confirmed by the injections.  There is no single value of $\alpha$
that gives the correct effective distance for all the injections; this is
consistent with the estimated systematic errors in the calibration.
Unfortunately the calibration line was not present during the time the
hardware injections were performed, so we cannot directly compare a measured
calibration with the result of the injections.

\section{Conclusions and Future Work}
\label{s:conclusions}
The analysis of the hardware injections in S1 was very productive. It allowed
us to test the software injections and check that the correct parameters are
recovered for injected signals. We recovered all the injections for the
signals between $10$ and $80$ kpc with a timing accuracy of $1/4096$ s. The
detected distance of the signals was correct to within the calibration
uncertainty of approximately $30\%$. We confirmed that the variation of
signal-to-noise ratio with calibration scaled as we expected.  In addition,
hardware injections were used to ensure that our pipeline did not veto real
signals due to using unsafe auxiliary channels as vetoes\cite{abbott2003b}.

The first science run lasted two weeks, so our hardware injections were
limited. The second LIGO science run (S2) took place 14 February---14 April
2003 and a more comprehensive set of hardware injections was performed with
the calibration line present. We are currently analyzing the data taken during
S2. The injection of inspiral signals into the interferometer will again form
an important part of our analysis.

\ack
We gratefully acknowledge the LIGO project and the LIGO Scientific
Collaboration, who made the first LIGO science run possible.
This work was supported by the National Science Foundation under
grants PHY-9970821 and PHY-0200852.

\end{document}